\UseRawInputEncoding
\documentclass[pra,superscriptaddress, notitlepage, reprint,twocolumn]{revtex4-1}
\usepackage[english]{babel}
\usepackage{amsmath,amsthm}
\usepackage{amsfonts}
\usepackage[pdfborder={0 0 0}, colorlinks=true, urlcolor=blue, linkcolor=blue, citecolor=blue]{hyperref}
\usepackage{color}
\usepackage{graphicx}
\usepackage{float}
\usepackage{subfigure}
\usepackage{lipsum}
\usepackage{epstopdf}
\usepackage{dcolumn}
\begin{document}

\title{Quantum estimation of Kerr nonlinearity in driven-dissipative systems}
\author{Dong  Xie}\email{xiedong@mail.ustc.edu.cn}

\affiliation{College of Science, Guilin University of Aerospace Technology, Guilin, Guangxi 541004, People's Republic of China}
\affiliation{State Key Laboratory for Mesoscopic Physics, School of Physics, Frontiers Science Center for Nano-Optoelectronics, and
Collaborative Innovation Center of Quantum Matter, Peking University, Beijing 100871, People's Republic of China}
\author{Chunling Xu}
\affiliation{College of Science, Guilin University of Aerospace Technology, Guilin, Guangxi 541004, People's Republic of China}
\author{An Min Wang}
\affiliation{Department of Modern Physics, University of Science and Technology of China, Hefei, Anhui 230026, People's Republic of China}

\begin{abstract}
We mainly investigate the quantum measurement of Kerr nonlinearity in the driven-dissipative system. Without the dissipation, the measurement precision of the nonlinearity parameter $\chi$ scales as ``super-Heisenberg scaling" $1/N^2$ with $N$ being the total average number of particles (photons) due to the nonlinear generator. Here,  we find that ``super-Heisenberg scaling" $1/N^{3/2}$ can also be obtained by choosing a proper interrogation time.  In the steady state, the ``super-Heisenberg scaling" $1/N^{3/2}$ can only be achieved when the nonlinearity parameter is close to 0 in the case of the single-photon loss and the one-photon driving or the two-photon driving. The ``super-Heisenberg scaling" disappears with the increase of the strength of the nonlinearity. When the system suffers from the two-photon loss in addition to the single-photon loss, the optimal measurement precision will not appear at the nonlinearity $\chi=0$ in the case of the one-photon driving. Counterintuitively, in the case of the two-photon driving we find that it is not the case that the higher the two-photon loss, the lower the measurement precision. It means that
the measurement precision of $\chi$ can be improved to some extent by increasing the two-photon loss.
\end{abstract}
\maketitle

\section{Introduction}
It is very important to improve the sensitivity and precision of parameters in modern science and technology\cite{lab1,lab2,lab3}.
The quantum mechanical effects, such as superposition and entanglement, offer a possibility to make the measurement sensitivity scale as $1/N$\cite{lab4,lab5}, which is the so-called Heisenberg scaling. Classical resources can only get the scaling $1/N$ at most, i.e., the standard quantum limit\cite{lab6}. However, by using nonlinear interaction or time-dependent evolutions, the scaling $1/N^{k}$ with $k>1$
can be obtained\cite{lab7,lab8,lab10,lab11,lab12,lab13,lab14,lab15,lab16,lab17,lab18,lab19}. It is dubbed as super-Heisenberg scalings\cite{lab7,lab10}, which is beyond the Heisenberg scaling. Ref.\cite{lab20,lab21} have some dispute about whether the super-Heisenberg scaling is really ¡°super-Heisenberg¡±. As shown in ref.\cite{lab19,lab20}, the super-Heisenberg scalings are actually Heisenberg from the perspective of the query complexity and all these different scalings can be seen as manifestations
of the Heisenberg uncertainty relations.  In order to better characterize the metrological advantages brought by the nonlinearity, we still define super-Heisenberg scaling as $\delta\chi\propto1/N^k$ with $k>1$.

Dissipation is inevitable in any physical systems. In recent years, the driven-dissipative systems have more and more theoretical and experimental attention. Strong optical nonlinearities have been obtained in different systems, such as cavity
quantum electrodynamics\cite{lab22}, optomechanical systems\cite{lab23}, Rydberg atomic systems\cite{lab24,lab25},  and superconducting
circuit QED systems\cite{lab26,lab27,lab28,lab29}. However, very little work has been done on the estimation of the nonlinearity itself in dissipative systems.
Recently, Jayakrishnan \textit{et.al.}\cite{lab30} utilized dissipatively coupled systems where the coupling is produced via
interaction with the vacuum of the electromagnetic field to enhance sensing of nonlinearities without utilizing a
commensurate gain-loss profile. They have showed that a small change in the nonlinearity can lead to a substantial change in the induced spin current. Strictly speaking, sensitivity is not equal to measurement precision. It is necessary to systematically explore the nonlinear estimation precision within the framework of quantum metrology.

In this article, we investigate the quantum estimation of the nonlinearity in the dissipative system. Without dissipation, the estimation precision of the nonlinearity can be scales as $1/N^2$ due to the nonlinear generator. Intuitively, dissipation will reduce measurement precision. Whether a super-Heisenberg scaling can still be obtained in the dissipative system is an interesting question. We try to solve this question and find that by choosing a proper interrogation time, the super-Heisenberg scaling $1/N^{3/2}$  can be achieved without extra driving.
Without initial state preparation, the nonlinearity can also be measured in the steady state from  the competition between the dissipation and the extra coherent one-photon driving or the two-photon driving. In the case of the one-photon loss, we find that the super-Heisenberg scaling $1/N^{3/2}$ can be obtained for the very weak nonlinearity  $\chi\rightarrow0$; the super-Heisenberg scaling disappears with the increase of the strength of the nonlinearity. Together with the two-photon loss, the optimal measurement precision does not appear when the nonlinearity $\chi$ is 0 in the case of the one-photon driving. Counterintuitively, in the case of the two-photon driving we find that it is not the case that the higher the two-photon
dissipation, the lower the measurement precision.

This article is organized as follows. In Section II, we introduce the dissipative Kerr model and the super-Heisenberg scaling of the nonlinearity is obtained by choosing a proper interrogation time. In Section III, we derive the measurement uncertainty of the nonlinearity in the steady state from the competition between the coherent one-photon driving and the single-photon loss. In Section IV, the measurement uncertainty of the nonlinearity is achieved in the case of the two-photon driving. In Section V,  we discuss about the effect of the two-photon loss on the quantum estimation. We make a brief conclusion in Section VI. The corresponding derivation is stated in Appendix A.

\section{dissipative Kerr nonlinear system}
We consider a nonlinear oscillator subject to the single-photon loss. The unitary dynamics can be dominated by  the Hamiltonian (from now on, $\hbar=1$ and the hats on the operators are omitted for simplifying the description)
\begin{align}
H=\omega_c a^\dagger a+\frac{\chi }{2} a^{\dagger2} a^2,
\label{eq:1}
\end{align}
where $a$ ($a^\dagger$) is the annihilation (creation) operator of the resonator, $\omega_c$ is the frequency of the resonator and $\chi$
 is the Kerr nonlinearity to be estimated.
Including the single-photon loss, the dynamic of the system is described by the Lindblad master equation\cite{lab31}
\begin{align}
\frac{d}{dt}\rho=\mathcal{L}\rho=-i[H,\rho]+\gamma\mathcal{D}[a]\rho,
\label{eq:2}
\end{align}
where $\rho$ is the density matrix, $\mathcal{L}$ is the Liouvillian super-operator, $\gamma$ is the one-photon decay rate, and the standard dissipator is defined as $\mathcal{D}[O]\rho=O\rho O^\dagger-[O^\dagger O\rho+\rho O^\dagger O]/2$.

Without extra driving, the steady state of the system will be in the vacuum state. Hence, the Kerr nonlinearity $\chi$ can not be estimated in the steady state. The optimal interrogation time should not be the characteristic time of reaching the steady state.

We assume the initial state is $(|0\rangle+|2N\rangle)/\sqrt{2}$, where $N$  is the total average
number of particles. Without the single-photon loss ($\gamma=0$), at time $t$ the state $|\psi(t)\rangle$ can be described by
\begin{align}
|\psi(t)\rangle=[e^{-i(2N^2-N)\chi t-i N \omega_c t}|2N\rangle+|0\rangle]/\sqrt{2}.
\label{eq:3}
\end{align}

Based on the root-mean-square error, the famous Cram\'{e}r-Rao (QCR) bound offers a good estimation limit of parameter precision
\begin{align}
\delta \chi\geq\frac{1}{\sqrt{\nu}\sqrt{F_Q}}\equiv \delta\chi_{\textmd{min}},
\label{eq:4}
\end{align}
where $\nu=T/t$ is the total number of repeated experiments with $T$ being the total experimental period, $F_Q$ is the quantum Fisher information (QFI).
For the pure state in Eq.~(\ref{eq:3}), the QFI can be calculated by
\begin{align}
F_Q=4[\langle\psi'(t)|\psi'(t)\rangle-|\langle\psi'(t)|\psi(t)\rangle|^2],
\label{eq:5}
\end{align}
where $|\psi'(t)\rangle=\frac{d}{d\chi}|\psi(t)\rangle$. As a result, we can achieve the best precision of the optimal measurement
\begin{align}
\delta\chi_{\textmd{min}}=\frac{1}{N(2N-1)\sqrt{Tt}}.
\label{eq:6}
\end{align}
For $N\gg1$, $\delta\chi_{\textmd{min}}\simeq\frac{1}{2N^2\sqrt{Tt}}$, which means that it is beyond the Heisenberg scaling ($\delta\chi_{\textmd{min}}\sim1/N$) in the framework of the linear metrology approach. Here, the nonlinearity can obtain the so-called super-Heisenberg scaling ($\delta\chi_{\textmd{min}}\sim1/N^2$).

When the system is subject to the single-photon loss, the state of the system will be a mixed state. Due to the weak symmetry of the system, the Liouvillian super-operator can be diagonalized by the formalism of the third quantization\cite{lab32,lab33,lab34}. We obtain the final density matrix based on the exact eigenvalues and eigenvectors of the Liouvillian superoperator given by  A. McDonald and A. A. Clerk\cite{lab35}. Given the same initial state $(|0\rangle+|2N\rangle)/\sqrt{2}$, the diagonalized state can be expressed as (see Appendix A)
\begin{align}
\rho(t)=\sum_{j=1}^{2N-1}\rho_{jj}|j\rangle\langle j|+\sum_{k=\pm}\lambda_k |\lambda_k\rangle\langle\lambda_k|.\label{eq:7}
\end{align}

\begin{figure}
\includegraphics[scale=0.65]{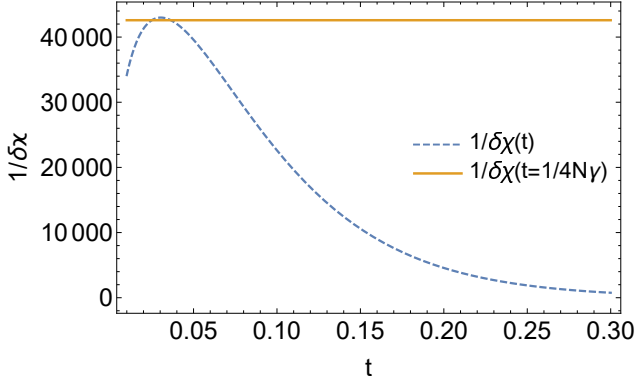}
\caption{\label{fig.1}The evolution diagram of reciprocal of the measurement uncertainty $1/\delta\chi$ over time. Here, the dimensionless parameters are set to be: $T=100$, $\gamma=0.1$ and $N=100$.}
\end{figure}

By calculating the QFI(see Appendix A),  we can obtain the measurement precision at the interrogation time $t=1/(4N\gamma)$, which are described by
\begin{align}
\delta \chi_{\textmd{min}}(t=\frac{1}{4N\gamma})=\sqrt{\frac{\sqrt{e}\gamma+e \gamma+e \gamma(1-e^{\frac{1}{4N}})^{2N} }{2N(2N-1)^2T}}.\label{eq:8}
\end{align}
As shown in Fig.~(\ref{fig.1}), we can see that $1/\delta \chi(t=\frac{1}{4N\gamma})$ is close to the maximum of $1/\delta \chi(t)$. It means that $1/(4N\gamma)$ approximates the optimal interrogation time.

For the large number of particles $N\gg1$, the measurement precision can be expressed as
\begin{align}
\delta \chi_{\textmd{min}}(t=\frac{1}{4N\gamma})\approx\sqrt{\frac{\sqrt{e}\gamma+e \gamma}{8N^3T}}.\label{eq:9}
\end{align}
From the above equation, we can see that the super-Heisenberg scaling ($1/N^{3/2}$) is achieved by choosing a prober interrogation time in the case of the single-photon loss. Comparing Eq.~(\ref{eq:6}) and Eq.~(\ref{eq:9}), we find that the scaling changes from $1/N^2$ to $1/N^{3/2}$.
It means that the measurement precision  is reduced by the single-photon loss. Still, it is possible to beat Heisenberg limit $1/N$ by choosing the right measurement time.

\section{quantum estimation with the coherent one-photon driving}
When one want to estimate the nonlinearity parameter $\chi$ in the steady state, extra driving is necessary. One of its advantages is that no advance preparation of states is required. In this section, we consider that there is a coherent one-photon driving, the corresponding Hamiltonian is given by
\begin{align}
H_{\textmd{op}}=i \Omega(a^\dagger e^{-i\omega_pt}-a e^{i\omega_pt}).\label{eq:10}
\end{align}
In the rotating frame, the total Hamiltonian is described by
\begin{align}
H_1=\Delta a^\dagger a+\frac{\chi}{2} a^{\dagger2} a^2+i \Omega(a^\dagger -a),\label{eq:11}
\end{align}
where
$\Delta=\omega_c-\omega_p$ denotes the detuning with $\omega_c$ and $\omega_p$ being the frequencies of the resonator and the driving field, respectively.
Subject to the single-photon loss, we can obtain the evolution of the system density matrix like Eq.~(\ref{eq:2}),
\begin{align}
\frac{d}{dt}\rho=\mathcal{L}\rho=-i[H_1,\rho]+\gamma\mathcal{D}[a]\rho.
\label{eq:12}
\end{align}
The analytical correlation functions of the above master equation in the steady-state can be achieved by the method of the complex P-representation\cite{lab36} and the Keldysh-Heisenberg equations\cite{lab37}, which are given by
\begin{align}
\langle a^{\dagger l}a^k\rangle=Tr[\rho_{ss}a^{\dagger l}a^k]\nonumber\\
=\frac{(\epsilon^*)^l\epsilon^k\Gamma(\beta^*)\Gamma(\beta)_0F_2(\beta^*+l,\beta+k;2|\epsilon|^2)}{\Gamma(\beta^*+l)\Gamma(\beta+k)_0F_2(\beta^*,\beta;2|\epsilon|^2)}.
\label{eq:13}
\end{align}
where $\Gamma(\beta)$ is the gamma special function and the generalized hypergeometric function $_0F_2(\beta^*,\beta;2|\epsilon|^2)=\sum_{m=0}^\infty \frac{\Gamma(\beta^*)\Gamma(\beta)(2|\epsilon|^2)^m}{\Gamma(\beta^*+m)\Gamma(\beta+m)m!}$ with $\epsilon=2\Omega/(i\chi)$ and $\beta=(2\Delta-i\gamma)/\chi$.

We consider a weak nonlinearity limit $\chi\rightarrow0$  first. Then, the steady state will be close to Gaussian state due to that  the effective Hamiltonian in Eq.~(\ref{eq:12}) is a quadratic form for $\chi\rightarrow0$\cite{lab38}.
The QFI for Gaussian state can be obtained through the fidelity by Pinel \textit{et. al.} in 2013\cite{lab39},
\begin{align}
F_{1Q}=\frac{1}{2(1+P_\theta^2)}Tr[(\mathcal{C}^{-1}_\theta\mathcal{C}^{'}_\theta)^2]+\frac{2P_\theta'^2}{1-P_\theta^4}\nonumber\\+{\langle\mathbf{X}^\top\rangle}'_\theta\mathcal{C}^{-1}_\theta\langle\mathbf{X}\rangle'_\theta,
\label{eq:14}
\end{align}
where $P_\theta=\frac{1}{2 d}$, $d=\sqrt{Det \mathcal{C}}$ and $A'_\theta$ is the term by term derivative of $A_\theta$ with respect to $\theta$. The quadrature operators are defined as $q :=\frac{1}{\sqrt{2}}(a+a^\dagger)$ and $p :=\frac{1}{i\sqrt{2}}(a-a^\dagger)$ with $a$ ($a^\dagger$) as the annihilation (creation) operator for a single bosonic mode. And the vector of quadrature operators is $\mathbf{X}=(q,p)^\top$. The entries of the covariance matrix $\mathcal{C}$ are defined as $\mathcal{C}_{ij}:=\frac{1}{2}\langle \{\mathbf{X}_i,\mathbf{X}_j\}\rangle-\langle \mathbf{X}_i\rangle\langle \mathbf{X}_j\rangle$, where $\langle\bullet\rangle=Tr[\bullet\rho_{ss}]$.
Substituting Eq.~(\ref{eq:13}) into Eq.~(\ref{eq:14}), we can obtain the QFI with $\chi\rightarrow0$
\begin{align}
F_{1Q}=\frac{16N^3}{4\Delta^2+\gamma^2},
\label{eq:15}
\end{align}
where the average photon number $N=\langle a^\dagger a\rangle=\frac{4\Omega^2}{4\Delta^2+\gamma^2}$.
The characteristic time to reach the steady state is given by $\tau\approx1/\gamma$. Hence, the measurement precision of the nonlinearity $\chi$ is obtained according to Eq.~(\ref{eq:4})
\begin{align}
\delta \chi_{\textmd{min}}|_{\chi\rightarrow0}=\frac{\sqrt{4\Delta^2+\gamma^2}}{4\sqrt{\gamma T}N^{3/2}}.
\label{eq:16}
\end{align}

The measurement precision can also be  derived by the error propagation formula
\begin{align}
\delta \chi=\sqrt{\frac{\langle M^2\rangle-\langle M\rangle^2}{\nu |\frac{d\langle M\rangle}{d\chi}|^2}},
\label{eq:17}
\end{align}
where $M$ denotes a measurement operator.
Generally, the uncertainty from the error propagation formula is equal or greater than that from the QFI in Eq.~(\ref{eq:4}),i.e., $\delta \chi\geq\delta \chi_{\textmd{min}}$. When they are equal, $\delta \chi=\delta \chi_{\textmd{min}}$, $M$ is the optimal measurement operator.

We find that when $M$ is the quadrature operator $p$, the uncertainty $\delta \chi|_{\chi\rightarrow0}=\delta \chi_{\textmd{min}}|_{\chi\rightarrow0}$. It shows that the quadrature operator $p$ is the optimal operator for the very weak nonlinearity $\chi\rightarrow0$.

From Eq.~(\ref{eq:15}), what's interesting is that the super-Heisenberg scaling $1/N^{3/2}$ is obtained which is the same as the case of no coherent driving. And then let's see if we can still get this scaling $1/N^{3/2}$ as the nonlinearity increases.
\begin{figure}
 \includegraphics[scale=0.65]{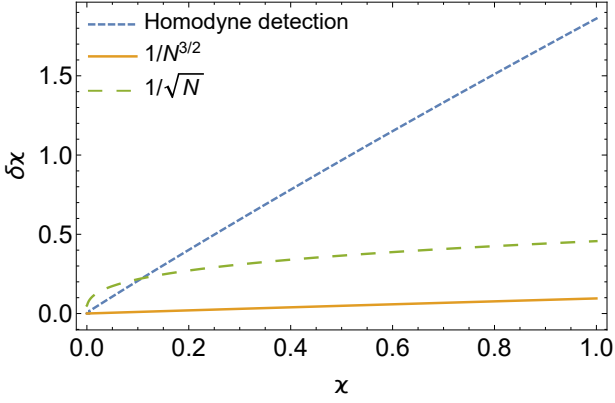}
  \caption{\label{fig.2}Diagram of the change of measurement uncertainty $\delta\chi$  with $\chi$. The information of $\chi$ is obtained by the Homodyne detection with the quadrature operator $p$. $1/N^{3/2}$ denotes the super-Heisenberg scaling. $1/\sqrt{N}$ denotes the quantum limit. Here, the dimensionless parameters are set to be: $\nu=1$, $\gamma=1$, $\Omega=100$, and $\Delta=0$.}
\end{figure}

As shown in Fig.~(\ref{fig.2}), the super-Heisenberg scaling can not obtained with the large value of $\chi$. Even the quantum limit can't be reached as $\chi$ increases. This shows that the single-photon loss has a worse effect on the measurement of large nonlinearity parameter $\chi$.

\section{quantum estimation with the two-photon driving}
In this section, we measure the nonlinearity parameter $\chi$ in the steady state with the two-photon driving, which is described by
\begin{align}
H_{\textmd{tp}}=\Lambda/2 (a^{\dagger2} e^{-i2\omega_pt}+a^2 e^{i2\omega_pt}).\label{eq:18}
\end{align}
The corresponding evolution of the system density matrix is described by
\begin{align}
\frac{d}{dt}\rho=\mathcal{L}\rho=-i[H_2,\rho]+\gamma\mathcal{D}[a]\rho.
\label{eq:19}
\end{align}
where the total Hamiltonian $H_2$ in the rotating frame is described by
\begin{align}
H_2=\Delta a^\dagger a+\frac{\Lambda}{2} (a^{\dagger2}+a^2),\label{eq:20}
\end{align}
The analytical correlation functions of the above master equation in the steady-state can also be derived by
the method of the complex P-representation and the Keldysh-Heisenberg equations,
\begin{align}
\langle a^{\dagger l}a^k\rangle
=\frac{1}{\mathcal{N}\sqrt{2^{l+k}}}\sum_{m=0}^\infty\frac{1}{m!}\mathcal{F}^*_{m+l}\mathcal{F}_{m+k},
\label{eq:21}
\end{align}
where $\mathcal{N}=\sum_{m=0}^\infty\frac{|\zeta|^{2m}}{m!}|_2F_1(-m,y,z;2)|^2$, $\mathcal{F}_{m}=(-\zeta)^m {_2F_1(-m,y,z;2)}$ and $_2F_1(-m,y,z;2)=\sum_{k=0}^\infty\frac{(-m)_k(y)_k 2^k}{(z)_k k!}$ denotes the generalized hypergeometric function with $(r)_k=\Gamma(r+k)/\Gamma(r)$, $\zeta=i\sqrt{2\Lambda/\chi}$, $y=(2\Delta-i\gamma)/(2\chi \zeta)$ and $z=(2\Delta-i\gamma)/\chi $.

For the nonlinearity $\chi\rightarrow0$, we can obtain the expectation values
\begin{align}
\langle a\rangle=\zeta/\sqrt{2},\ \langle a^2\rangle=\zeta^2/2,\
N=\langle a^\dagger a\rangle=\frac{2\Lambda}{\chi}.
\label{eq:22}
\end{align}
Substituting the above results and $M=p$ into Eq.~(\ref{eq:17}), the measurement uncertainty of $\chi$ is achieved
\begin{align}
\delta\chi=\sqrt{\frac{\chi^3}{\Lambda}}=\frac{2\sqrt{2}\Lambda}{N^{3/2}}.
\label{eq:23}
\end{align}
\begin{figure}
 \includegraphics[scale=0.6]{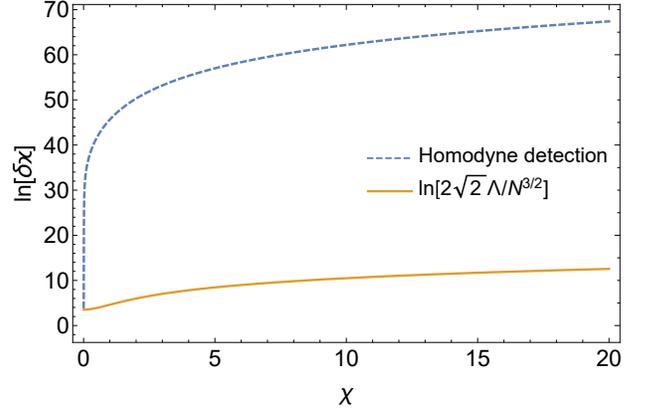}
  \caption{\label{fig.3}Diagram of the change of measurement uncertainty $\ln[\delta\chi]$  with $\chi$. The information of $\chi$ is obtained by the Homodyne detection with the quadrature operator $p$. $1/N^{3/2}$ denotes the super-Heisenberg scaling.  Here, the dimensionless parameters are set to be: $\nu=1$, $\Lambda=0.1$, $\gamma=1$, and $\Delta=0$.}
\end{figure}
This shows that the uncertainty $\delta \chi$ will be close to 0 when the nonlinearity $\chi$ is close to 0. It is due to that the average photon number will be close to infinity with the two-photon driving for $\chi\rightarrow 0$. As $\chi$ increases, the energy level difference also increases, leading to that it leaves fewer resources to measure the nonlinearity $\chi$, i.e., the system is hard to drive to high energy levels.

As shown in Fig.~(\ref{fig.3}), we  can see that the super-Heisenberg scaling quickly disappears as $\chi$ increases. It  is consistent with the previous case of the coherent single-photon driving.
\section{Subject to the two-photon loss}
Besides the single-photon loss, there are two-photon loss. In this section, we study the influence of the two-photon loss on the estimation of the nonlinearity.

The evolution of the system density matrix is described by \cite{lab39}
\begin{align}
\frac{d}{dt}\rho=\mathcal{L}\rho=-i[H_3,\rho]+\gamma\mathcal{D}[a]\rho+\kappa\mathcal{D}[a^2]\rho,
\label{eq:24}
\end{align}
where $H_3$ represents a general Hamiltonian including the coherent one- and two-photon driving
\begin{align}
H_3=\Delta a^\dagger a+\frac{\Lambda}{2} (a^{\dagger2}+a^2)+i \Omega(a^\dagger -a)\label{eq:25}.
\end{align}
The analytical correlation function in the steady-state can be also given by Eq.~(\ref{eq:21}), which just needs the corresponding parameter to be replaced by
\begin{align}
\zeta\rightarrow i\sqrt{2\Lambda/(\chi-i \kappa)},
z\rightarrow(2\Delta-i\gamma)/(\chi-i \kappa )\\
y\rightarrow[-i2\sqrt{2}\Omega+\zeta(\Delta-i\gamma)]/[2\zeta (\chi-i\kappa)].
\end{align}

\begin{figure}
 \includegraphics[scale=0.6]{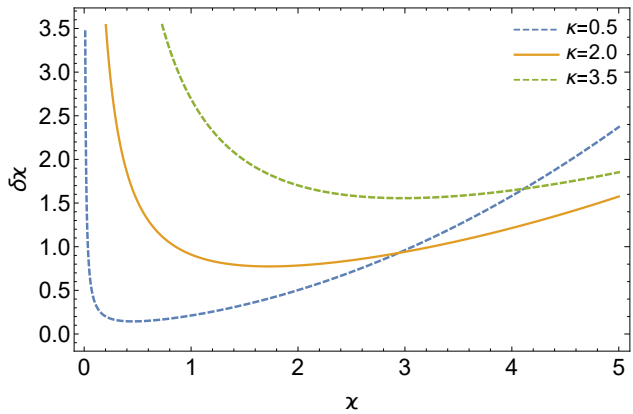}
  \caption{\label{fig.4}Diagram of the change of measurement uncertainty $\delta\chi$  with $\chi$. The information of $\chi$ is obtained by the Homodyne detection with the quadrature operator $p$.  Here, the dimensionless parameters are set to be: $\nu=1$, $\Lambda=0$, $\Omega=100$, $\gamma=10$, and $\Delta=0$.}
\end{figure}

When there is only the single-photon driving, i.e., $\Lambda=0$, the two-photon loss will reduce the measurement precision of $\chi$ as shown in Fig.~(\ref{fig.4}). What's more, the optimal measurement precision do not appear at the value $\chi=0$, which is very different from the situation where there is no two-photon loss. This means that the nonlinear two-photon loss shifts the value of $\chi$ corresponding to the optimal measurement precision. Intuitively, it is  because $\chi$ is replaced by $\chi-i \kappa $.

\begin{figure}
 \includegraphics[scale=0.62]{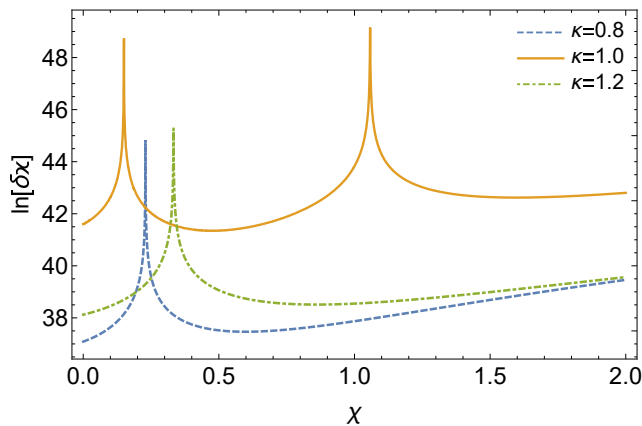}
  \caption{\label{fig.5}Diagram of the change of measurement uncertainty $\ln[\delta\chi]$  with $\chi$. The information of $\chi$ is obtained by the Homodyne detection with the quadrature operator $p$.  Here, the dimensionless parameters are set to be: $\nu=1$, $\Lambda=0.5$, $\Omega=0$, $\gamma=10$, and $\Delta=0$.}
\end{figure}

 When there is only the two-photon driving ($\Omega=0$), we can see that the optimal measurement precision do not appear at $\chi=0$ for some values of $\kappa$, as shown in Fig.~(\ref{fig.5}). Counterintuitively, what we found is that it's not that the higher the value of $\kappa$, the lower the measurement precision of $\delta \chi$ by comparing the line $``\kappa=1"$ and the line $``\kappa=1.2"$ in Fig.~(\ref{fig.5}). This means that the measurement precision of $\chi$ can be improved to some extent by increasing the two-photon dissipation in the case of the two-photon driving.

\section{conclusion}
We have investigated the quantum estimation of the nonlinearity in the Kerr system subject to the single-photon loss. Without the extra driving, the super-Heisenberg scaling $1/N^{3/2}$ can be obtained by using a proper interrogation time, which is shown to be close to the optimal interrogation time. With the coherent one-photon driving or the two-photon driving, we show that the super-Heisenberg scaling $1/N^{3/2}$ is only achieved when the nonlinearity $\chi=0$. In general, the nonlinearity  is very weak. Therefor, the coherent one-photon driving and the two-photon driving are useful in enhancing the measurement precision of the nonlinearity. With the
increase of the strength of the nonlinearity, the super-Heisenberg scaling or even the quantum limit can not be reached.
 Finally, we investigate the influence of the two-photon loss on the measurement precision of the nonlinearity. As a result, the optimal measurement
precision will appear at the nonlinearity $\chi\neq0$ in the case of the one-photon driving. Counter-intuitively, in the case of two-photon driving we find that it is not that the higher the two-photon dissipation, the lower the measurement precision. It means that the measurement precision of $\chi$ can be improved to some extent by increasing the two-photon dissipation.
It is worth noting that increasing the driving strength will improve the measurement precision of the nonlinearity. But whether it will increase the utilization of resources is a question worth further study.

The driven-dissipative model can be realizable in the present experiment. The one- and two-photon driven-dissipative resonator can be built in different platforms, such as, coupled two superconducting cavities through a Josephson junction\cite{lab40}, superconducting circuits QED\cite{lab41,lab42}, and optical Josephson interferometer\cite{lab43}.

\section*{Acknowledgements}
This research was supported by the National Natural Science Foundation of China under Grant No. 62001134, Guangxi Natural Science Foundation under Grant No. 2020GXNSFAA159047 and National Key R\&D Program of China under Grant No. 2018YFB1601402-2.

\section*{Appendix A}
In the third-quantized form, the inner product is $\langle B|A\rangle=Tr [B^\dagger A]$ and the annihilation and creation super-operator satisfy $a_L|\rho\rangle=|a\rho\rangle$, $a_R|\rho\rangle=|\rho a\rangle$, $a_L^\dagger|\rho\rangle=|a^\dagger\rho\rangle$, and $a_R^\dagger|\rho\rangle=|\rho a^\dagger\rangle$.

In the case of the single-photon loss, the propagator can be written as\cite{lab35}
\begin{align}
e^{\mathcal{L} t}=\sum_{m=-\infty}^\infty\sum_{\mu=0}^\infty e^{\lambda_{m,\mu}t}|r_{m,u}\rangle\langle l_{m,u}|,
\tag{S1}\label{eq:S1}
\end{align}
where $|r_{m,u}\rangle$ ($\langle l_{m,u}|$) are the right (left) eigenvectors of the superoperator $\mathcal{L}$,

$|r_{m,u}\rangle=$\[
 \frac{1}{\sqrt{\mu!(\mu+|m|)!}} \left\{
\begin{array}{ll}
(c_{+,m}^\dagger c_{-,m})^\mu (c_{+,m}^\dagger)^m|0^r_m\rangle,\ \
m\geq0\\
(c_{+,m}^\dagger c_{-,m})^\mu (-c_{-,m})^{-m}|0^r_m\rangle, m<0
  \end{array}
\right .\tag{S2}\label{eq:S2}\]
$ \langle l_{m,u}|=$
\[
\frac{1}{\sqrt{\mu!(\mu+|m|)!}} \left\{
\begin{array}{ll}
\langle0^r_m|(d_{+,m})^m(-d^\dagger_{-,m}d_{+,m})^\mu, m\geq0\\
\langle0^r_m|(d^\dagger_{-,m})^{-m}(-d^\dagger_{-,m}d_{+,m})^\mu, m<0
  \end{array}
\right.\tag{S3}\label{eq:S3}\]
And the corresponding eigenvalues are
\begin{align}
\lambda_{m,\mu}=-i(\omega_c-\chi)m-(\gamma+i\chi m)(|m|+2\mu)/2.\tag{S4}
\label{eq:S4}
\end{align}
Here, $d_{+,m}$ and $c_{-,m}$ ( $d_{-,m}^\dagger$ and $c_{+,m}^\dagger$) are linear combinations of $a_L$ and $a_R$ ($a_L^\dagger$ and $a_R^\dagger$) respectively, which are given by
\begin{align}
d_{+,m}=x_m a_L,
c_{-,m}=y_ma_L-z_ma_R,\tag{S5}\\
d_{-,m}^\dagger=x_m a_R^\dagger,
c_{+,m}^\dagger=z_ma_L^\dagger-y_ma_R^\dagger,\tag{S6}
\label{eq:S4}
\end{align}
where $x_m=\frac{im\chi +2\gamma}{\sqrt{2}(im\chi +\gamma)}$, $y_m=\frac{2\gamma}{\sqrt{2}(im\chi +2\gamma)}$ and $z_m=\frac{2im\chi +2\gamma}{\sqrt{2}(im\chi +2\gamma)}$. The right and left ``vacuum" states are Gaussian states
\begin{align}
0^r_m=|0\rangle\langle0|,\tag{S7}\\
0^l_m=\sum_{n=0}^\infty(\frac{\gamma}{\gamma+i m\chi})^n|n\rangle\langle n|.\tag{S8}
\label{eq:S8}
\end{align}

The density matrix at time $t$ can be achieved by the superoperator $\mathcal{L}$,
\begin{align}
|\rho(t)\rangle=e^{\mathcal{L} t}|\rho(t=0)\rangle.\tag{S9}
\label{eq:S9}
\end{align}
Using Eq.~(\ref{eq:S1}-\ref{eq:S9}) and the initial state $\rho(t=0)=(|0\rangle+|2N\rangle)(\langle0|+\langle2N|)/2$, we can obtain the analytical form of the density matrix $\rho(t)=\rho_{jk}|j\rangle\langle k|$ for $j,k=0,...,2N$, which is composed by
\begin{align}
&\rho_{00}=\frac{1}{2}+\frac{1}{2}(1-e^{-\gamma t})^{2N},\tag{S10}\\
&\rho_{(2N)0}=\rho_{0(2N)}^*=\frac{1}{2}e^{\lambda_{m=2N,\mu}},\tag{S11}\\
&\rho_{jj}=\sum_{\mu=j}^{2N}\frac{(2N)!e^{-\gamma \mu t}}{2(2N-\mu)!(\mu-j)!j!(-1)^{(\mu+j)}},\nonumber\\
&2N\geq j\geq1.\tag{S12}
\label{eq:S12}
\end{align}
All other matrix elements in the density matrix are 0.

The density matrix can be diagonalized
\begin{align}
\rho(t)=\sum_{j=1}^{2N-1}\rho_{jj}|j\rangle\langle j|+\sum_{k=\pm}\lambda_k |\lambda_k\rangle\langle\lambda_k|,\tag{S13}\label{eq:S13}
\end{align}
where the eigenvalues are $\lambda_\pm=\frac{\rho_{00}+\rho_{(2N)(2N)}}{2}\pm\sqrt{|\rho_{(2N)0}|^2+(\frac{\rho_{00}-\rho_{(2N)(2N)}}{2})^2}$ and the corresponding eigenvectors are
\begin{align}
 |\lambda_\pm\rangle=\frac{1}{\sqrt{\mathcal{N}_\pm}}(\rho_{0(2N)}|0\rangle+(\lambda_\pm-\rho_{00})|2N\rangle),\tag{S14}\label{eq:S14}
\end{align}
where the normalization factors are given by $\mathcal{N}_\pm={|\rho_{(2N)0}|^2+(\lambda_\pm-\rho_{00})^2}$.
Due to that the term $\sum_{j=1}^{2N-1}\rho_{jj}|j\rangle\langle j|$ and $\lambda_\pm$ are independent of the nonlinearity parameter $\chi$,
the QFI can be derived by
\begin{align}
F_{sQ}=\sum_{k=\pm}4\lambda_{k}\langle\lambda'_{k}|\lambda'_{k}\rangle-\sum_{j=\pm,k=\pm}\frac{8\lambda_{j}\lambda_{k}}{\lambda_{j}+\lambda_{k}}
|\langle\lambda_{j}|\lambda'_{k}\rangle|^2.\tag{S15}\label{eq:S15}
\end{align}

Substituting Eq.~(\ref{eq:S13}-\ref{eq:S14}) into Eq.~(\ref{eq:S15}), we can obtain the analytical form of the QFI
\begin{align}
F_{sQ}=\sum_{k=\pm}\frac{4\lambda_k|\rho_{(2N)0}|^2(2N-4N^2)^2t^2}{\mathcal{N}_k}(1-\frac{|\rho_{(2N)0}|^2}{\mathcal{N}_k})\nonumber\\
-\frac{16(\rho_{00}\rho_{(2N)(2N)}-|\rho_{0(2N)}|^2)|\rho_{(2N)0}|^4(2N-4N^2)^2t^2}{\mathcal{N}_+\mathcal{N}_-}.\tag{S16}\label{eq:S16}
\end{align}
The measurement uncertainty of $\chi$ at time $t$ is given by
\begin{align}
\delta ^2\chi_{\textmd{min}}(t)=\frac{t}{TF_{sQ}}.\tag{S17}\label{eq:S17}
\end{align}

The optimal interrogation time $t_{op}$ can be derived by
\begin{align}
\frac{d\delta \chi_{\textmd{min}}}{dt}|_{t_{op}}=0.\tag{S18}\label{eq:S18}
\end{align}
For $N\gg1$, we can obtain the approximate optimal interrogation time $t_{op}\approx \frac{1}{4N\gamma}$.
Based on the optimized interrogation time, the measurement uncertainty $\delta \chi_{\textmd{min}}(t=\frac{1}{4N\gamma})$ is achieved
\begin{align}
\delta \chi_{\textmd{min}}(t=\frac{1}{4N\gamma})=\sqrt{\frac{\sqrt{e}\gamma+e \gamma+e \gamma(1-e^{\frac{1}{4N}})^{2N} }{2N(2N-1)^2T}}.\tag{S19}\label{eq:S19}
\end{align}

\end{document}